\documentclass[conference]{IEEEtran}
\IEEEoverridecommandlockouts
\usepackage{cite}
\DeclareUnicodeCharacter{2061}{}
\usepackage{lettrine}
\usepackage{amsmath,amssymb,amsfonts}
\usepackage{algorithmic}
\usepackage{graphicx}
\usepackage{textcomp}
\usepackage{caption}
\usepackage{subcaption}
\newcommand\norm[1]{\left\lVert#1\right\rVert}
\usepackage{xcolor}
\def\BibTeX{{\rm B\kern-.05em{\sc i\kern-.025em b}\kern-.08em
    T\kern-.1667em\lower.7ex\hbox{E}\kern-.125emX}}

\usepackage{mathtools}

\DeclarePairedDelimiter\floor{\lfloor}{\rfloor}
\usepackage{steinmetz}
\usepackage{authblk}

\usepackage{array}

\usepackage[ruled,norelsize]{algorithm2e}
\makeatletter
\newcommand{\removelatexerror}{\let\@latex@error\@gobble}
\makeatother

\usepackage{caption}
\usepackage{subcaption}
\usepackage{blindtext}
\usepackage{threeparttable}
\usepackage{array}
\begin{document}

\title{Pre-scaling and Codebook Design for Joint Radar and Communication Based on Index Modulation}

\author{Shengyang Chen, Aryan Kaushik, \IEEEmembership{Member,~IEEE,} and Christos Masouros, \IEEEmembership{Senior Member,~IEEE}
\thanks{Shengyang Chen and Christos Masouros are associated with the Department of Electronic and Electrical Engineering, University College London, London U.K. (e-mail: \{shengyang.chen.20, c.masouros@ucl.ac.uk\}).}
\thanks{Aryan Kaushik is associated with the School of Engineering and Informatics, University of Sussex, Brighton, U.K. (e-mail: a.kaushik@ucl.ac.uk).}
\thanks{This work was supported by the UK Engineering and Physical Sciences Research Council (EPSRC) Grant number EP/S026657/1, and the UK MOD University Defence Research Collaboration (UDRC) in Signal Processing.}
\thanks{This paper has been developed on master's thesis of Shengyang Chen.}
}

\maketitle

\begin{abstract}
   This paper develops an efficient index modulation (IM) approach for the joint radar-communication (JRC) system  based on a multi-carrier multiple-input multiple-output (MIMO) radar. The communication information is embedded into the transmitted radar pulses by selecting the corresponding indices of the carrier frequencies and antenna allocations, providing two degrees of freedom. Our contribution involves the development of a novel codebook based minimum Euclidean distance (MED) maximization and a constellation randomization pre-scaling (CRPS) scheme for efficient IM-JRC transmission. It can be inferred that the IM approach integrating the CRPS scheme followed by the codebook design maximizes the signal-to-noise ratio gain. The numerical results support the effectiveness of the proposed approach and show enhanced bit error rate performance when compared to the existing baseline. 
\end{abstract}

\begin{IEEEkeywords}
Joint radar and communication, index modulation, constellation randomization, frequency/spatial agility. 
\end{IEEEkeywords}

\section{Introduction}
\lettrine{\textbf{T}}{he} frequency competition between the communication and radar systems will inevitably cause severe congestion within the finite spectrum of sub-6 GHz. To save RF resources, the sub-6 GHz spectrum allocated to radar systems can be made available for sharing between radar and wireless systems \cite{griffithsPROC2015}. Going beyond spectral co-existence, the joint radar-communication (JRC) systems integrate the complete hardware platform and the transmitted signal waveform of the two systems \cite{liuTCOM2020}. Recently the JRC systems have been advanced for hardware efficient architectures and employing low resolution converters \cite{aryanICC2021, dizdarICC2021} that are sought for in emerging fifth generation (5G) wireless communications \cite{aryanTGCN2021}. 

The JRC systems are classified into three categories \cite{zhangJSTSP2021}. One is \emph{communication-centric} JRC which implements radar sensing as the secondary function of the existing communication system, e.g., \cite{strum2011}. 
Second is \emph{radar-centric} JRC which integrates wireless communication as the secondary function of the existing radar system, e.g., \cite{eldarTSP2020}. 
Third is \emph{joint JRC} that offers tunable trade-off between both the operations, e.g., \cite{aryanICC2021}. The design concept of the radar-centric JRC systems is to embed the communication information into the transmitted radar pulses and meanwhile maintain the sensing performance \cite{hanIET2013}. Furthermore, \cite{hassanienSPM2019} classifies the information embedding approaches for the radar-centric JRC systems into beam pattern modulation, index modulation (IM), and fast-time modulation. 

The IM approach which utilizes the randomness and uniqueness of the radar system to convey the information outperforms the conventional methods with enhanced data throughput and robust error performance \cite{eldarTSP2020}. It aims to generate a comprehensive coding mechanism so that the blocks of information bits can be mapped accordingly to the indices of function blocks (communication codewords) within the JRC system \cite{clancyWCNC2015}. Example function blocks can cover the spatial position of the antennas in the multiple-input multiple-output (MIMO) system, the carrier frequency in the multiple-carriers system, and the type of waveform in the multi-waveform system. The work in \cite{eldarTSP2020_2} extends frequency agile radar \cite{axelsson2007} into the multi-carrier MIMO transmission. 
\cite{eldarTSP2020} further extends this model for the JRC systems by exploiting the frequency and antenna agility. However, the exploitation of information embedding with more effective signal processing approaches and enhanced error performance has not been widely studied.

\emph{Contributions}: This paper implements an efficient IM based information embedding approach in JRC systems that utilizes the spectral and spatial randomness of the radar pulse signals. For that, the communication codewords are embedded in the transmitted radar pulses by selecting the corresponding indices of the carrier frequencies and antenna allocations. Firstly, we propose a novel codebook based minimum Euclidean distance (MED) maximization approach to select the best combinations of antennas and frequencies to form our signal constellation. We then further augment performance through a constellation randomization pre-scaling (CRPS) scheme for dual functionality transmission. We show the benefits of integrating the CRPS scheme followed by the codebook design that maximizes the signal-to-noise ratio (SNR) gain. Our numerical evaluation reveals an enhanced bit error rate (BER) for the proposed approaches when compared to the existing baseline.


\vspace{-2mm}
\section{System Model}
\label{sec:system}
We consider IM-JRC, which is implemented based on a multi-carrier MIMO radar (MCMR) \cite{eldarTSP2020_2} through IM. In MCMR, the collection of available carrier frequencies is as
$ F=\{f_\textrm{c}+m\Delta f | m\in (0,M-1)\}$,  
where $M$ is the number of carrier frequencies available at the transmitter (TX), $f_\textrm{c}$ is the initial carrier frequency, and $\Delta f$ is the frequency step. Let $L_\textrm{R}$ be the number of the TX antennas, with $d=10 \frac{c}{f_\textrm{c}}$ being the spacing between the neighboring antenna elements and $c$ 
is the speed of light. Suppose that there are $N$ radar pulses repeatedly transmitted during each coherent processing interval of MCMR. Thus, the radar pulses will be transmitted within the time instance from $n T_\textrm{r}$ to $n T_\textrm{r}+T_\textrm{p}$, where $n \in (0,N-1)$.

During operation, MCMR randomly chooses a subset of $K$ carrier frequencies from $F$ for the $n$-th radar pulse as
 $F_\textrm{n}= \{\Omega_\textrm{n,k}  | k\in(0,K-1)\}$, 
where $F_\textrm{n}\subset F$, and $\Omega_\textrm{n,k}$ is an individual selected carrier frequency. Subsequently, these frequencies are randomly allocated among the phased array antennas, so that each antenna utilizes a single carrier frequency from $F_\textrm{n}$. Let $f_\textrm{n,l}$ be the carrier frequency transmitted by the $l$-th antenna. If the same antenna element also happens to transmit at $\Omega_\textrm{n,k}$, then $\Omega_\textrm{n,k}=f_\textrm{n,l}$. In MCMR, the TX radar waveform is
\begin{equation}
   	\phi(f,t)= \text{rect}(t/T_\textrm{p})\text{exp}(j2\pi ft). 
\end{equation}
In addition, the radar signal emitted from each antenna element should be multiplied with a weighting function to direct the antenna beam to the optimum angle $\theta$ \cite{eldarTSP2020}, given by
\begin{equation}
w_\textrm{l}(\theta)= \text{exp}(j2\pi f_\textrm{c} ld\text{sin}\theta/c).
\end{equation}

To formulate the transmitted signal vector $\mathbf{x}(n,t)\in \mathbb{C}^{L_\textrm{R}}$ of the entire antenna array for the $n$-th radar pulse, define $\mathbf{x}_\textrm{k}(n,t)$ as the subset of $\mathbf{x}(n,t)$ that transmits at frequency $\Omega_\textrm{n,k}$. Then the finalized transmission model of MCMR is 
\begin{equation}\label{eq:tx_mcmr}
\mathbf{x}(n,t) \!\!=\!\! \sum_{k=0}^{K-1}\mathbf{x}_\textrm{k}(n,t)\!\!=\!\!\sum_{k=0}^{K-1}\mathbf{P}(n,k)\mathbf{w}(\theta)\phi(\Omega_\textrm{n,k},t\!-\!nT_\textrm{r}), 
\end{equation}
where $\mathbf{P}(n,k)\in \{0,1\}^{L_\textrm{R}\times L_\textrm{R}}$ is the diagonal matrix containing the spatial agility information. It depends on the diagonal vector $\mathbf{p}(n,k)\in \{0,1\}^{L_\textrm{R}}$, whose $l$-th element is $1$ if $f_\textrm{n,l}=\Omega_\textrm{n,k}$, and is $0$ otherwise. 
The parameter $\phi(\Omega_\textrm{n,k},t-nT_\textrm{r})$ denotes the transmitted waveform at time instance $t$, which carries the frequency agility information.


For IM-JRC, considering a single pulse transmission, i.e., $N=1$, we simplify notations as $\mathbf{x}(t)=\mathbf{x}(n,t),\mathbf{x}_\textrm{k}(t)=\mathbf{x}_\textrm{k} (n,t),\mathbf{P}_\textrm{k}=\mathbf{P}(n,k)$ and $\mathbf{w}_\textrm{k}=\mathbf{w}(\theta)$. At the initial stage of transmission, IM-JRC chooses a subset of $K$ carrier frequencies out of $F$. The set of all possible frequency selections $\zeta$ is 
$\zeta=\{F^{(i)} | F^{(i)} \subset F,|F^{(i)}| =K,\,\, i=0,1,2,...\}$, 
where $F^{(i)}$ is an individual selection of frequencies. The cardinality of $\zeta$ is
$|\zeta|=\frac{M!}{K!(M-K)!}$. 
We define $L_\textrm{K}=L_\textrm{R}/K$, where $L_\textrm{K}$ is assumed to be an integer greater than $1$, i.e., each carrier frequency is transmitted by precisely $L_\textrm{K}$ antenna elements. 
$P$ is the collection of all potential antenna allocations, given by
$P\!\!=\!\!\{\mathbf{P}_\textrm{k}^{(i)}, k\in(0,K\!-\!1) | \text{trace}(\mathbf{P}_\textrm{k}^{(i)})\!\!=\!L_\textrm{K}, \,\,i\!=\!0,1,2,..\}$,
where $\{\mathbf{P}_0^{(i)},...,\mathbf{P}_\textrm{{K-1}}^{(i)}\}$ represents one possible allocation pattern. The cardinality of $P$ is
$|P|=\frac{L_\textrm{R}!}{{(L_\textrm{K}!)}^K}$.

IM-JRC integrating the frequency and spatial index modulation
and following \cite{duhamel2010}, the maximum number of bits that can be embedded in a single radar pulse transmission is
\begin{equation}
   \text{log}_2 |\zeta||P| = \text{log}_2 \frac{M!}{K!(M-K)!} + \text{log}_2 \frac{L_\textrm{R}!}{{(L_\textrm{K}!)}^K}.
\end{equation}
Assuming $L_\textrm{C}$ receiver (RX) antenna elements, we have
\begin{equation}
\mathbf{y}_\textrm{C} (t)=\mathbf{H}\mathbf{x}(t)+\mathbf{n}_\textrm{C} (t)=\sum_{k=0}^{K-1} \mathbf{H}\mathbf{x}_\textrm{k} (t)+\mathbf{n}_\textrm{C}(t), 
\end{equation}
where $\mathbf{y}_\textrm{C}(t)\in \mathbb{C}^{L_\textrm{C}}$ represents the received JRC signal. $\mathbf{H} \in \mathbb{C}^{L_\textrm{C} \times L_\textrm{R}}$ is the Rayleigh flat fading channel. 
$\mathbf{n}_\textrm{C}(t)\in \mathbb{C}^{L_\textrm{C}} $ is the additive white Gaussian noise (AWGN).

Before the sampling process, the received JRC signal $\mathbf{y}_\textrm{C}(t)$ is firstly down-converted by a factor of $e^{-j2\pi f_\textrm{c} t}$ to remove the high frequency component $f_\textrm{c}$. After that, IM-JRC employs the Nyquist sampling theorem and samples $\mathbf{y}_\textrm{C}(t)$ at the time instances of $iT_\textrm{s}$, where $T_\textrm{s}$ is defined as the sampling period and $i\in(0,\floor{T_\textrm{p}/T_\textrm{s}}$). Assuming the input bitstream is randomly and equally distributed, 
the optimum Nyquist sampling period is given by $T_\textrm{s}=1/M\Delta f$.
Considering $L_\textrm{T}=\floor{T_\textrm{p}/T_\textrm{s}}+1$ as the number of sampled outputs per pulse and $c_\textrm{k}\in(0,M-1)$ being the frequency index corresponding to the selected carrier frequencies $\Omega_\textrm{k}$ which is
$c_\textrm{k}= (\Omega_\textrm{k}-f_\textrm{c})/\Delta f$. 
Recall \eqref{eq:tx_mcmr}, and after sampling the finalized reception model of IM-JRC is 
\begin{equation}
\mathbf{Y}_\textrm{C}=\sum_{k=0}^{K-1} \mathbf{H}\mathbf{P}_\textrm{k} \mathbf{w}_\textrm{k} \boldsymbol{\Phi}_{c_\textrm{k}}^T+\mathbf{N}_\textrm{C} ,   
\end{equation}
where $\mathbf{Y}_\textrm{C}\in \mathbb{C}^{L_\textrm{C} \times L_\textrm{T}}$ and $\mathbf{N}_\textrm{C}\in\mathbb{C}^{L_\textrm{C} \times L_\textrm{T}}$ are the sampled RX signal, 
and sampled AWGN signal, respectively. $\boldsymbol{\Phi}_{c_\textrm{k}}\in \mathbb{C}^{L_\textrm{T}}$ is the down-converted and sampled transmission waveform, computed as
$\boldsymbol{\Phi}_{c_\textrm{k}}={[1,e^{j2\pi c_\textrm{k} \Delta f T_\textrm{s}\times 1},...,e^{j2\pi c_\textrm{k} \Delta f T_\textrm{s}\times(L_\textrm{T}-1)}]}^T$.
It is also regarded as the baseband signal carrying the information of the frequency index $c_\textrm{k}$ of a single radar pulse. 

\section{Proposed Codebook Design and CRPS Schemes}
In this section, we introduce the novel codebook design and CRPS schemes for IM-JRC.
Following \cite{eldarTSP2020}, for IM-JRC the maximum-likelihood (ML) decoder shows best BER performance in terms of estimating the frequency indices $\{c_\textrm{k}\}$ and antenna allocation patterns $\{\mathbf{P}_\textrm{k}\}$, which are previously embedded in the radar pulses through IM at the TX. The ML detection for IM-JRC is expressed as
\begin{equation}\label{eq:ml_estimation}
{\{{c}_\textrm{k},{\mathbf{P}}_\textrm{k}\}}_{k=0}^{K-1}
={_{\{{c}_\textrm{k},{\mathbf{P}}_\textrm{k}\}}^{\text{arg min}} \norm{\mathbf{Y}_\textrm{C}-\sum_{k=0}^{K-1} \mathbf{H}\mathbf{P}_\textrm{k}\mathbf{w}_k\boldsymbol{\Phi}_{c_\textrm{k}}^T}_F^2} ,
\end{equation}
where $\norm{.}_F$ denotes the Frobenius norm. Additionally, we assume that the RX has \emph{a priori} knowledge of $K$, the desired pointing angle $\theta$, and perfect channel state information (CSI). 
 
\subsection{Codebook Design}
When the available codewords, i.e., combinations of antenna and frequency indices, are not equal to a power of $2$ \cite{duhamel2010}, IM-JRC only utilizes a subset of the available codewords for transmission. Defining the subset of codewords as valid codewords $V$, then
$|V|=2^B \le |\zeta||P|$,
where $B$ is the maximum number of bits that can be conveyed in an individual radar pulse. In general, the codebook design approach aims to locate the optimum $|V|$ valid codewords.
To define “optimum”, 
\cite{yangTWC2016} pointed out that the BER performance of the communication system is highly dependent on the MED between the neighbouring constellation symbols, i.e., valid codewords, where the MED is formulated as 
\begin{equation}
\textrm{MED}=d_\textrm{min}=_{i\neq j}^\text{min}\norm{\mathbf{Y}_\textrm{i}-\mathbf{Y}_\textrm{j}}_F^2. 
\end{equation}
Consequently, the designed codebook for IM-JRC should comprise the subset of valid codewords with maximum MED. 

 \begin{figure}[t]
 \centering 
 \includegraphics[width=0.4\textwidth, trim=36 100 55 0,clip]{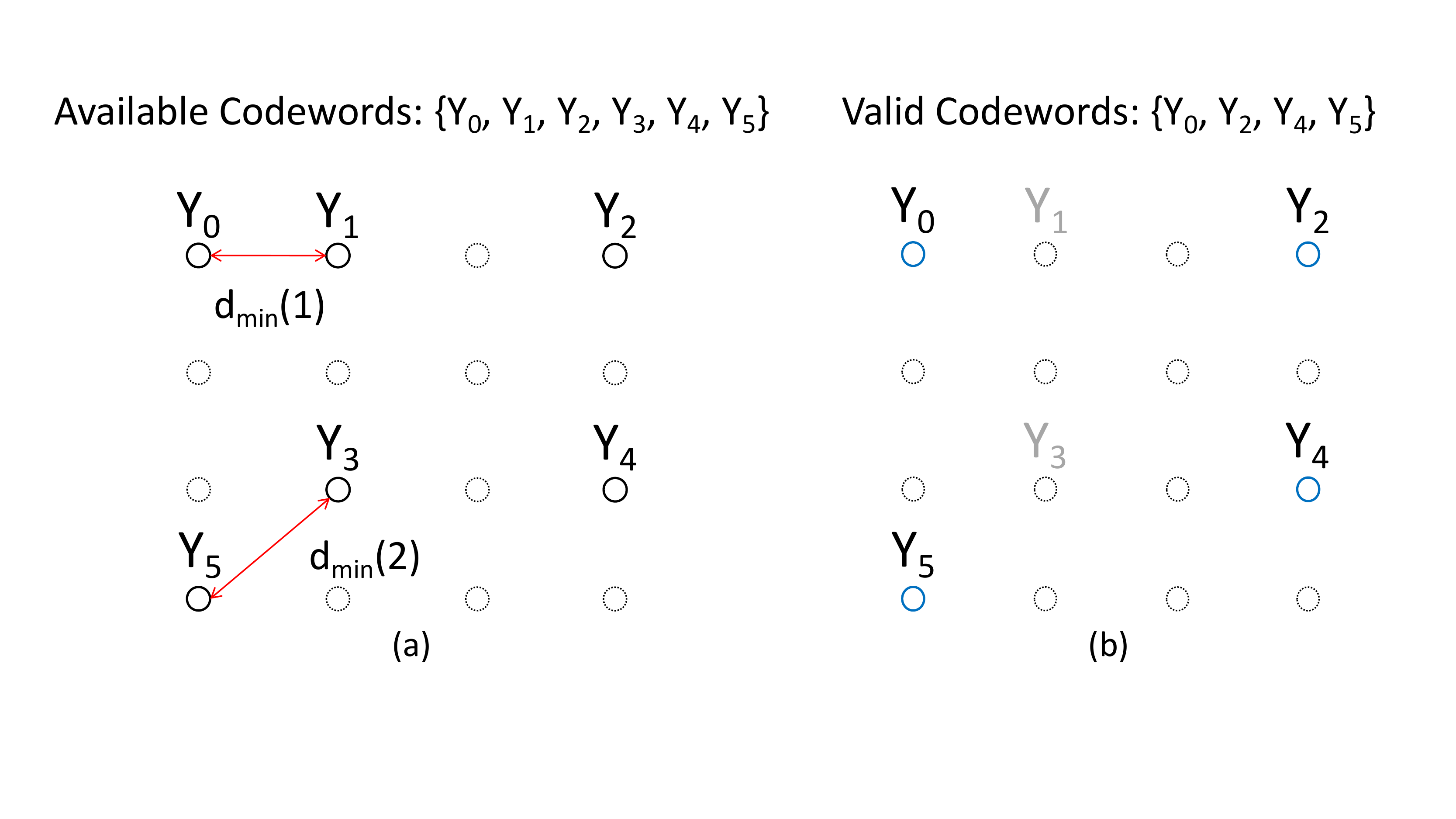}
 		\caption{Illustration of Codebook Design Approach: (a) Original Codewords; (b) Designed Codebook.}
 		\vspace{-6mm}
 \end{figure}

Fig. 1 demonstrates the proposed codebook design approach. In Fig. 1(a), there are $C=|\zeta||P|=6$ codewords available for transmission $\{\textrm{Y}_\textrm{0},...,\textrm{Y}_\textrm{5}\}$.
Hence, IM-JRC can convey a maximum of $\floor{\text{log}_2(C)}=2$ bits information with four valid codewords. The design philosophy is to locate the “worst” two codewords and make them idle. In Step 1, IM-JRC computes for the current MED and the corresponding neighbouring codewords, which are $\textrm{Y}_\textrm{0}$ and $\textrm{Y}_\textrm{1}$. After that, the system compares their second-minimum Euclidean distances with the remaining codewords. In this example, $d_{\textrm{Y}_0-\textrm{Y}_3}> d_{\textrm{Y}_1-\textrm{Y}_3}$; thus, the system keeps $\textrm{Y}_\textrm{0}$ and removes $\textrm{Y}_\textrm{1}$. In Step 2, IM-JRC repeats the procedure in Step 1, and it decides to keep $\textrm{Y}_\textrm{5}$ and make $\textrm{Y}_\textrm{3}$ idle. The remaining four valid codewords $\{\textrm{Y}_\textrm{0},\textrm{Y}_\textrm{2},\textrm{Y}_\textrm{4},\textrm{Y}_\textrm{5}\}$ comprise the finalized codebook design. The designed codebook maximizes the MED between the codewords, thereby reducing the probability of detection errors and enhancing the BER performance. 

\vspace{-1mm}
\subsection{Constellation Randomization Pre-Scaling (CRPS)}
A constellation randomization approach is presented in \cite{masourosTVT2016, masourosCL2014} to improve the performance of spatial modulation in the MIMO communication systems by maximizing the system MED. 
The operating principle is to randomize the original constellation diagram by multiplying codewords with a range of randomly generated transmit pre-scaling (TPS) factors. 
We extend that technique for IM-JRC systems, termed as constellation randomization pre-scaling (CRPS). 

Let $\mathbf{X}_\textrm{C}$ represent the original constellation diagram. At the beginning of CRPS, a number of normalized TPS factors are generated, given by $\boldsymbol{\alpha} = \{\boldsymbol{\alpha}_\textrm{d}|d \in (1,D)\}$, where $\boldsymbol{\alpha}$ is the set containing all the TPS factors with the cardinality of $D$. The parameter $\boldsymbol{\alpha}_\textrm{d} \in \mathbb{C}^{L_\textrm{R}}$ is one particular TPS factor, where each entry is randomized from the standard normal distribution, and $\mathbb{E}[\mathbf{\alpha}_\textrm{d}]=1$ where $\mathbb{E}[.]$ is the expectation operator. Defining $\mathbf{A}_\textrm{d} \in \mathbb{C}^{L_\textrm{R}\times L_\textrm{R}}$ as the diagonal matrix of $\boldsymbol{\alpha}_\textrm{d}$. Then the original constellation diagram is randomized by multiplying it with each of the TPS matrices, such as 
$\mathbf{X}_\textrm{C-CRPS}\!\!=\!\!\mathbf{A}_\textrm{d} \mathbf{X}_\textrm{C}\!\!=\!\!\sum_{k=0}^{K-1} \mathbf{A}_\textrm{d} \mathbf{P}_\textrm{k} \mathbf{w}_\textrm{k} \boldsymbol{\Phi}_{c_\textrm{k}}^T$,
which is the collection of the randomized constellation diagrams. As shown in Fig. 2, the system MED varies accordingly with the randomization of the constellation diagram. Consequently, there are possibilities for the new MEDs to be greater than the original MED, where the possibilities increase proportionally with the number of TPS factors. Later, IM-JRC determines the optimal TPS factor which results in the maximum MED among the “pre-scaled” constellation symbols, hence the optimum BER performance, for which
\vspace{-4mm}
\begin{align}
\{\boldsymbol{\alpha}_\textrm{o},\mathbf{A}_\textrm{o}\}&\!\!=\!\!\text{arg} \max_d {_{\{c_{k_1},\mathbf{P}_{k_1}\}\neq\{c_{k_2},\mathbf{P}_{k_2}\}}^{\text{min}}}||\sum_{k=0}^{K-1} \mathbf{A}_\textrm{d} \mathbf{P}_{k_1} \mathbf{w}_\textrm{k} \boldsymbol{\Phi}_{c_{k_1}}^T \nonumber\\
&-\sum_{k=0}^{K-1}\mathbf{A}_\textrm{d}\mathbf{P}_{k_2} \mathbf{w}_\textrm{k} \boldsymbol{\Phi}_{c_{k_2}}^T||_F^2,
\end{align}
where $\{\boldsymbol{\alpha}_\textrm{o},\mathbf{A}_\textrm{o}\}$ are the optimal TPS factor and matrix, respectively. As shown in \cite{masourosTVT2016}, information of $\{\boldsymbol{\alpha}_\textrm{o},\mathbf{A}_\textrm{o}\}$ can be made available to the RX, or the RX can independently identify the TPS given an agreed codebook and CSI. Accordingly, the ML decoder in \eqref{eq:ml_estimation} is modified as
\begin{equation}
{\{{c}_\textrm{k},{\mathbf{P}}_\textrm{k}\}}_{k=0}^{K-1}={_{\{{c}_\textrm{k},{\mathbf{P}}_\textrm{k}\}}^{\text{arg min}} \norm{\mathbf{Y}_\textrm{C}-\sum_{k=0}^{K-1} \mathbf{H}\mathbf{A}_o \mathbf{P}_\textrm{k} \mathbf{w}_k\boldsymbol{\Phi}_{c_\textrm{k}}^T}_F^2}.
\end{equation}
   
 \begin{figure}[t]
 \centering 
 \includegraphics[width=0.4\textwidth, trim=36 85 55 0,clip]{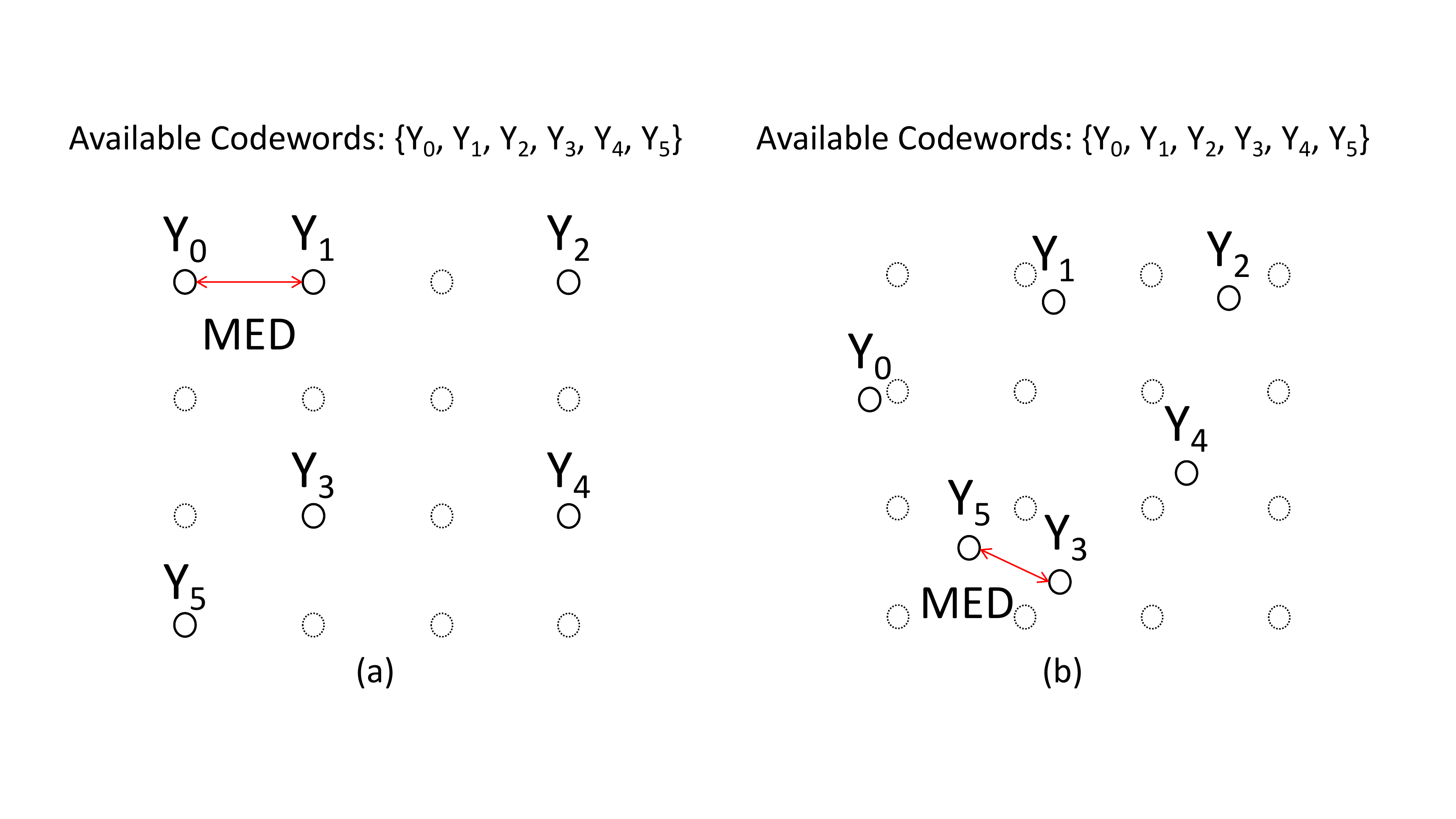}
 		\caption{Illustration of CRPS: (a) Original Constellation Symbols; 
(b) Randomized Constellation Symbols.}
 		\vspace{-2mm}
 \end{figure}

\begin{table}
    \centering
\def\arraystretch{1}
\begin{tabular}{ |c|c|}
 \hline
 \textbf{Method} & \textbf{Complexity order} \\
 \hline
 IM-Codebook design & $[K L_\textrm{R}+(2K-1)L_\textrm{T}]L_\textrm{R} |\zeta||P|$ \\ & $+\frac{3L_\textrm{R} L_\textrm{T}+1}{2} \sum_{i=1}^Q(|\zeta||P|-i+1)(|\zeta||P|-i)$ 
	\\ & $+[(L_\textrm{R}+3)L_\textrm{C} L_\textrm{T}+1](|\zeta||P|-Q)N$ \\
 \hline
 IM-CRPS approach & $[K L_\textrm{R}+(2K-1)L_\textrm{T}]L_\textrm{R} |\zeta||P|+[[L_\textrm{R}^2 L_\textrm{T}$\\ &$+((3L_\textrm{R}  L_\textrm{T}+1)(|\zeta||P|-1))/2]|\zeta||P|+1]D$\\&$+[(L_\textrm{R}+3)L_\textrm{C} L_\textrm{T}+1]|\zeta||P|N$ \\
 \hline
\end{tabular}
\caption{Computational complexity of the proposed schemes.}
\label{tab:Table1}
\vspace{-6mm}
\end{table}

 \begin{figure}[t]
 \centering 
 \includegraphics[width=0.45\textwidth, trim=50 0 0 0,clip]{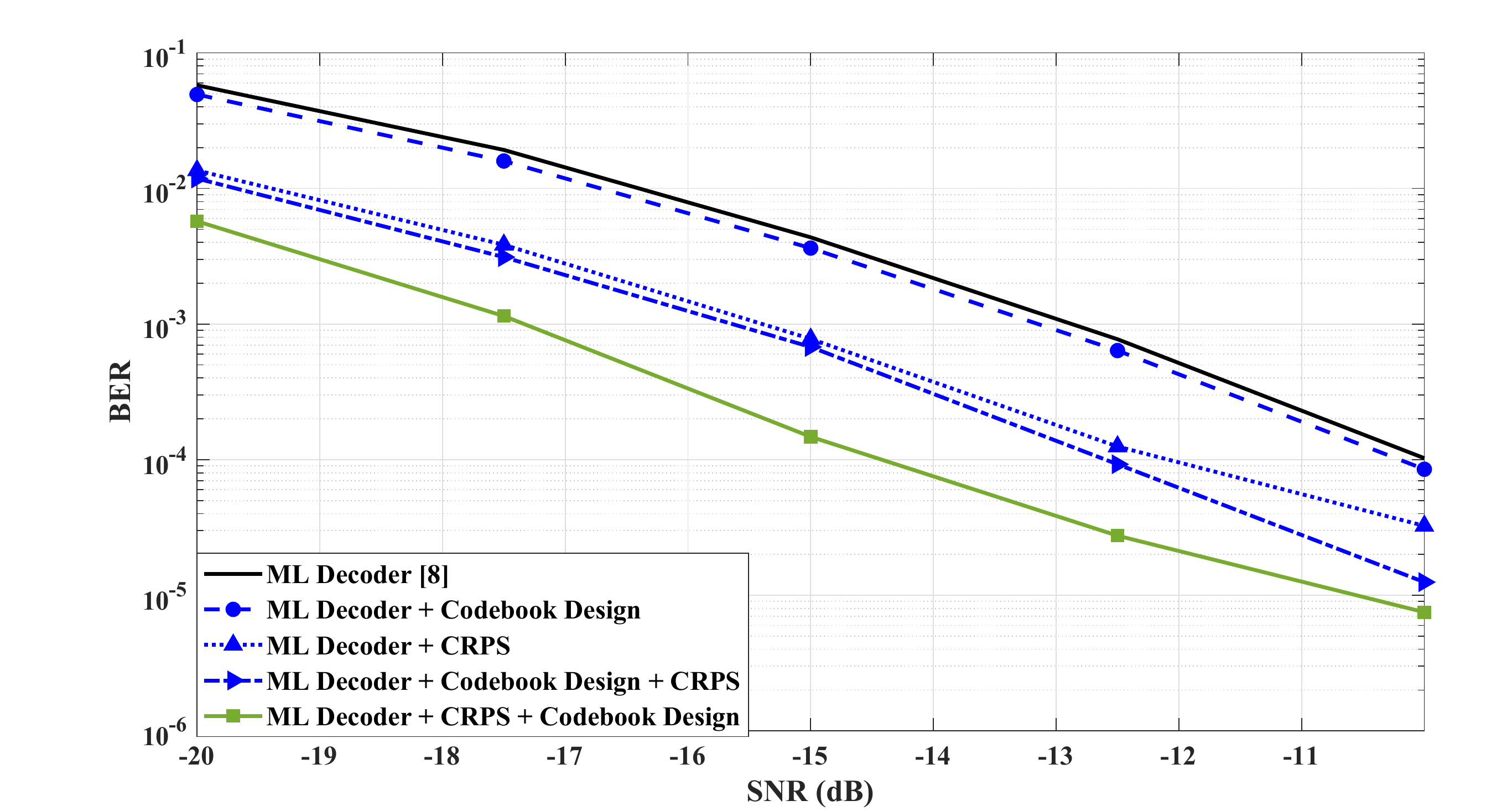}
 		\caption{BER performance of the proposed codebook design and CRPS schemes for $M=7$ and $L_\textrm{R}=6$.}
 		\vspace{-5mm}
 \end{figure}

The overall computational complexity of IM with codebook design and CRPS scheme is provided in Table 1, where $Q$ is the number of codewords to be eliminated, expressed as $Q=|\zeta||P|-|V|=|\zeta||P|-2^B$ and $B=\floor{\text{log}_2|⁡\zeta||P|}$. The complexity of the CRPS scheme also depends on $D$, which is the cardinality of the randomly generated TPS factors. 

\section{Simulation Results}
\label{sec:simulation}
We set the system parameters as, unless stated otherwise, $M$ = 7, $f_\textrm{c}$ = 1.9 GHz, $\Delta f$ = 10 MHz, $L_\textrm{R}$ = 6, $\theta$ = $\frac{\pi}{4}$, $c = 3\times 10^8$ m/s, $T_\textrm{r}$ = 2 $\mu s$, $T_\textrm{p}$ = 1 $\mu s$, $K$ = 2, $L_\textrm{C}$ = 4, and $D$ = 100. Under this condition, the JRC model can convey a maximum of $8$-bits information during each radar pulse transmission. Additionally, the transmit signal power is normalized to unity. The BER results are averaged over $N=1\times10^5$ radar pulses.

Fig. 3 shows the BER performance of the proposed schemes and compare it with ML decoder baseline \cite{eldarTSP2020}. 
The codebook design scheme slightly outperforms the ML decoder as it eliminates the idle communication codewords in the constellation diagram. 
The CRPS scheme outperforms the codebook design approach as we can observe that this scheme meets the BER levels of $1\times10^{-3}$ and $1\times10^{-4}$ at the SNRs of -15.4 and -12.1 dB, giving the SNR gains of 2.5 and 2.1 dB, respectively.

Furthermore, in the case of codebook design followed by CRPS, the codebook design firstly eliminates 164 idle codewords and enlarges the temporal MED; after that, the CRPS scheme randomizes the remaining 256 valid codewords and maximizes the average symbol distance. However, this randomization inevitably decreases the system MED. For the case of CRPS followed by codebook design, the CRPS scheme randomizes the overall 420 communication codewords and maximizes the average symbol distance while leaving a set of lower symbol distances. In the following, these undesirable distances are handled by the codebook design approach during the elimination of the idle codewords. As a result, the finalized constellation diagram is more suitable for the communication transmission, which also accounts for the optimal BER performance, e.g., the corresponding SNR gains at the BER levels of  $1\times10^{-3}$ and $1\times10^{-4}$ are both 4.4 dB.


The data rate being the maximum number of bits embedded per radar pulse, firstly we investigate in Fig. 4 the impact of varying $M$ on the BER performance of the optimal integrated CRPS and codebook design scheme. Following the Nyquist sampling period $T_\textrm{s}=\frac{1}{M \Delta f}$, it can be observed that when $M$ increases, the sampling period narrows, and the reliability of the decoding process is enhanced accordingly, which accounts for the BER performance enhancement of the IM-JRC model. This result also supports the trade-off between data rate and BER performance. We can also observe that the optimal proposed scheme outperforms ML decoder baseline for all the cases. Similarly Fig. 4 observes the BER performance with varying number of TX antennas $L_\textrm{R}$. The optimal integrated CRPS and codebook design approach again outperforms the ML decoder baseline for different $L_\textrm{R}$ values.

 \begin{figure}[t]
 \centering 
 \includegraphics[width=0.455\textwidth, trim=40 0 0 0,clip]{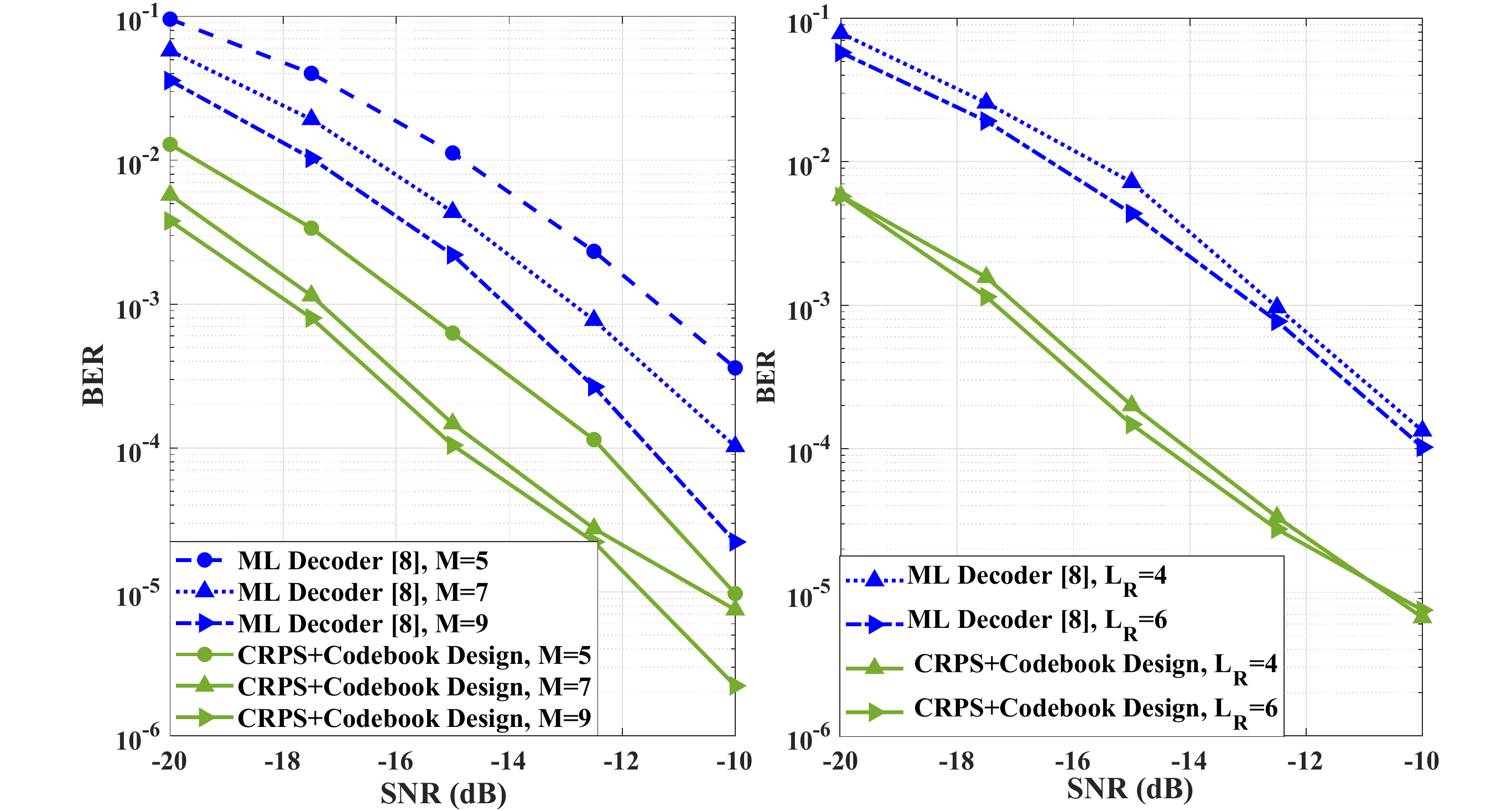}
 		\caption{BER performance of the integrated CRPS and codebook design for varying $M$ ($L_\textrm{R}=6$) and varying $L_\textrm{R}$ ($M=7$).}
 		\vspace{-5mm}
 \end{figure}

\section{Conclusion}
This paper proposes the novel and efficient codebook based MED maximization and CRPS schemes to employ IM in JRC systems with spectral and spatial agility. The proposed approaches outperform the state-of-the-art ML decoder baseline, and the integration of the CRPS scheme followed by the codebook design leads to the best optimal solution in terms of BER performance. In the specific case, employing CRPS followed by codebook design outputs the maximum SNR gain of around 4.4 dB when compared to the baseline approach.


\end{document}